\documentclass[12pt]{article}
\usepackage{latexsym,amsfonts,amssymb}
\makeatletter
\@addtoreset{equation}{subsection}
\makeatother

\begin{document}
\begin{center}
{\Large \bf The Regular Universe}
\\[1.5cm]
 {\bf Vladimir S.~MASHKEVICH}\footnote {E-mail:
  Vladimir.Mashkevich100@qc.cuny.edu}
\\[1.4cm] {\it Physics Department
 \\ Queens College\\ The City University of New York\\
 65-30 Kissena Boulevard\\ Flushing\\ New York
 11367-1519} \\[1.4cm] \vskip 1cm

{\large \bf Abstract}
\end{center}
A regular (i.e., singularity-free) cycling cosmological model is advanced. In the model, there are only two constants: the gravitational constant (or the Planck time) and the cosmic period. The radius of the universe is a simple periodic function of cosmic time. The regularity of the construction is achieved via the condition that the measure associated with metric be the Haar measure on the 3-space. The possibility of the construction is due to dark matter---as long as it is treated not as a particle matter, but as a compensational tensor field. The metrodynamical equation, i.e., an equation for metric, is derived and expressions for the pressure, energy and momentum compensons are obtained.

\newpage
\section*{Introduction}

If a physical theory contains singularities, this means that the theory may be incomplete. As long as the source of a metric field in General Relativity is treated as an ordinary matter, this results in singularities of metric and of the matter distribution---both local (black holes) and global, or cosmic (the Big Bang). It appears natural to try to get rid of the singularities, i.e., to construct a complete, or regular theory of the universe.

The regularity of the universe means above all the regularity of the spacetime manifold. The latter is a pair $(M^{4},g)$ [1] where $M^{4}$ is a connected four-dimensional Hausdorff $C^{\infty}$ manifold and $g$ is a Lorentz metric on $M^{4}$. The Lorentz metric is regular, i.e., nonsingular, which means that its determinant is not equal to zero. However, the determinant depends on a coordinate system; therefore, using the determinant directly, it is impossible to formulate the nonsingularity condition geometrically, i.e., in a coordinate-free way, and to incorporate the condition into metric dynamics.

A coordinate-free formulation of the nonsingularity condition may be achieved by means of the measure associated with metric. Metric of the universe induces the Riemannian metric on the 3-sphere, which may be represented in the form of $R^{2}(t)h_{ij}(t,\underline{x})\mathrm{d}x^{i}\mathrm{d}x^{j}$ with a preassigned radius of the universe $R(t)$.

The nonsingularity condition: The measure associated with the metric $h_{ij}$ is the Haar measure [2] on the 3-sphere. It is this condition that provides for the absence of singularities.

A possibility of such a construction arises in connection with dark matter---if it is treated not as a particle matter [3], but rather as a compensational tensor field, or the compenson [4]. A pivotal idea of the construction is the following. Since the compenson is not a particle field, there exists no equation for the compenson proper. An equation in addition to the Einstein equations should be introduced for metric. The additional equation is the result of the nonsingularity condition and it provides for the absence of singularities.

The resulting construction describes the regular, i.e., singularity-free cycling universe.

As to the absence of black holes, it should be pointed out that a contracting-expanding (or pulsating) massive star may simulate the behavior of a black hole: an external observer will see the star reach the horizon only after an infinitely long time [5]. A pure compenson object may appear to be a black hole as well.

A compensating term in the dynamical (i.e., spatial) components of the extended Einstein equation is represented in the simplest form---as a pressure compenson. This makes it possible to eliminate the term from the equation, which results in the metrodynamical equation---a dynamical equation for metric. The Haar measure condition may be regarded as an integral of motion of the metrodynamical equation.

The pressure, energy and momentum compensons are represented in an explicit form.

\section{Compenson}

\subsection{The extended Einstein equation. Compenson}

The extended Einstein equation,
\begin{equation} 
G^{\nu}_{\mu}=8\pi\varkappa(T_{\mu}^{\nu}+\Theta_{\mu}^{\nu})\,,\quad \mu,\nu=0,1,2,3
\end{equation}
includes apart from the matter energy-momentum tensor
\begin{equation} 
T_{\mu}^{\nu}=(\Psi,\hat{T}_{\mu}^{\nu}\Psi)
\end{equation}
the compenson, i.e., a compensational tensor field $\Theta_{\mu}^{\nu}$. In (1.1.1), $G^{\nu}_{\mu}$ is the Einstein tensor and $\varkappa=t_{\mathrm{Planck}}^{2}$ is the gravitational constant $(c=\hbar=1)$. The state vector $\Psi$ is guided by the Schr\"{o}dinger equation.

\subsection{Synchronous gauge}

Spacetime manifold is
\begin{equation} 
M^{4}=T\times S^{3}
\end{equation}
where $T$ is cosmic time and the 3-sphere $S^{3}$ is cosmic space. Correspondingly, metric is of the form
\begin{equation} 
\mathrm{d}s^{2}=\mathrm{d}t^{2}+g_{ij}\mathrm{d}x^{i}\mathrm{d}x^{j}\,,\quad i,j=1,2,3
\end{equation}
(synchronous [5], or temporal [6] gauge). Here $(x^{1},x^{2},x^{3})$ are dimensionless coordinates on $S^{3}$ and the dimensions are these:
\begin{equation} 
[g_{ij}]=[t^{2}]=T^{2}
\end{equation}

\subsection{Stress, energy and momentum compensons}
Any tensor splits into the spatial, energy and momentum components:
\begin{equation} 
A^{\nu}_{\mu}=(A^{j}_{i},A^{0}_{0},A_{i}^{0})
\end{equation}
and equation (1.1.1) splits accordingly:
\begin{equation} 
G^{j}_{i}=8\pi\varkappa(T_{i}^{j}+\Theta_{i}^{j})
\end{equation}
\begin{equation} 
G^{0}_{0}=8\pi\varkappa(T_{0}^{0}+\Theta_{0}^{0})
\end{equation}
\begin{equation} 
G^{0}_{i}=8\pi\varkappa(T_{i}^{0}+\Theta_{i}^{0})
\end{equation}
Here $\Theta^{j}_{i},\,\Theta^{0}_{0}$ and $\Theta^{0}_{i}$ are the stress, energy and momentum compensons, respectively.

Now, (1.3.2) is  a system of dynamical equations, and (1.3.3), (1.3.4) determine the energy-momentum compenson:
\begin{equation} 
\Theta_{\mu}^{0}=\frac{1}{8\pi\varkappa}G_{\mu}^{0}-T_{\mu}^{0}
\end{equation}

\subsection{The cosmological constant, dark energy, and dark matter}

Let us establish the relation of our treatment with the conventional one. Rewrite (1.1.1) in the form
\begin{equation} 
G^{\nu}_{\mu}-\Lambda g^{\nu}_{\mu}=8\pi\varkappa(T^{\nu}_{\mu}+
T_{\mathrm{dark\,matter}}{}^{\nu}_{\mu})
\end{equation}
where $\Lambda$ is the cosmological constant and
\begin{equation} 
T_{\mathrm{dark\,matter}}{}^{\nu}_{\mu}:=\Theta_{\mu}^{\nu}-\frac{\Lambda}{8\pi\varkappa}
g^{\nu}_{\mu}
\end{equation}
is the dark matter energy-momentum tensor.

Next,
\begin{equation} 
G^{0}_{0}=8\pi\varkappa(\varrho_{\mathrm{matter}}+\varrho_{\mathrm{dark\,matter}}+
\varrho_{\mathrm{dark\,energy}})
\end{equation}
where
\begin{equation} 
\varrho_{\mathrm{matter}}=T_{0}^{0}
\end{equation}
\begin{equation} 
\varrho_{\mathrm{dark\,matter}}=T_{\mathrm{dark\,matter}}{}^{0}_{0}=
\Theta_{0}^{0}-\frac{\Lambda}{8\pi\varkappa}
\end{equation}
\begin{equation} 
\varrho_{\mathrm{dark\,energy}}=\frac{\Lambda}{8\pi\varkappa}
\end{equation}

\section{Radius of the universe, the Haar measure condition, and dynamical equations}

\subsection{Radius introduced}

Put
\begin{equation} 
g_{ij}(t,\underline{x})=-R^{2}(t)h_{ij}(t,\underline{x})\,,\quad \underline{x}:=(x^{1},x^{2},x^{3})\in S^{3}
\end{equation}
with the dimensions
\begin{equation} 
[R]=T,\quad [h_{ij}]=T^{0}
\end{equation}
so that
\begin{equation} 
\mathrm{d}s^{2}=\mathrm{d}t^{2}-R^{2}(t)h_{ij}(t,\underline{x})
\mathrm{d}x^{i}\mathrm{d}x^{j}
\end{equation}
Here $R(t)$ is the radius of the universe, which is subject to specification, and $h_{ij}$ is the Riemannian metric on $S^{3}$, which is determined by dynamical equations.

\subsection{Stress compenson}

The compenson is not a particle field, therefore there exist no equations for the compenson proper. The energy-momentum compenson is determined by (1.3.3), (1.3.4). But the stress compenson $\Theta_{i}^{j}$ is contained in dynamical equations (1.3.2), so that it appears that a natural approach is to introduce additional equations for metric. To minimize the number of the equations, $\Theta_{i}^{j}$ should be reduced to a single scalar function $\vartheta$.

Put
\begin{equation} 
\Theta_{i}^{j}=\Theta_{i}^{j}[\vartheta]
\end{equation}

\subsection{The Haar measure condition}

In (1.3.2), there are 6 equations for 7 quantities $h_{ij},\;\vartheta$, so that it is necessary to introduce one more equation for metric $h_{ij}$.

Introduce a condition on the measure associated with the metric $h_{ij}$,
\begin{equation} 
\mathrm{d\mu}_{t}(\underline{x})=\sqrt{|h|}\mathrm{d}\underline{x}=
\sqrt{|h|}\mathrm{d}x^{1}\mathrm{d}x^{2}\mathrm{d}x^{3}\,,
\quad |h|=\mathrm{det}(h_{ij})
\end{equation}
The condition is this: (2.3.1) is the Haar measure [3] on $S^{3}$. The latter is associated with the metric [8]
\begin{equation} 
\mathrm{d}\chi^{2}+\sin^{2}\chi(\mathrm{d}\theta^{2}+\sin^{2}\theta\,\mathrm{d}\varphi^{2})\,,
\quad 0\leq \chi\leq\pi,\;0\leq\theta\leq\pi,\;0\leq\varphi\leq 2\pi
\end{equation}
and is
\begin{equation} 
\mathrm{d}\mu(\omega)=
\sin^{2}\chi\sin\theta\,\mathrm{d}\chi\mathrm{d}\theta\mathrm{d}\varphi\,,
\quad\omega=(\chi,\theta,\varphi)
\end{equation}
Thus, the condition is that (2.3.1) coincide with (2.3.3):
\begin{equation} 
\mathrm{d\mu}_{t}(\underline{x})=\mathrm{d}\mu(\omega)
\end{equation}
hence
\begin{equation} 
\frac{\partial}{\partial t}|h|=0
\end{equation}

Next, it holds that [8]
\begin{equation} 
\frac{\partial}{\partial t}\ln|h|=h^{ij}\dot{h}_{ij}\,,
\quad ^{.}=\frac{\partial}{\partial t}
\end{equation}
thus,
\begin{equation} 
h^{ij}\dot{h}_{ij}=0
\end{equation}
Equations (1.3.2) for $h_{ij}$ are of the second order in $t$, so we pass on to the equation
\begin{equation} 
\frac{\partial}{\partial t}(h^{ij}\dot{h}_{ij})=0
\end{equation}
i.e.,
\begin{equation} 
h^{ij}\ddot{h}_{ij}+\dot{h}^{ij}\dot{h}_{ij}=0
\end{equation}
Now (2.3.4) and (2.3.7) are constraints.

Note that equations (2.3.9), (2.3.4), (2.3.7) are 3-covariant (coordinate-free).

\subsection{Dynamical equations}

Dynamical equations for metric $h_{ij}$ and the stress compenson function $\vartheta$ are
\begin{equation} 
G_{i}^{j}-8\pi\varkappa\Theta_{i}^{j}[\vartheta]
=8\pi\varkappa T_{i}^{j}
\end{equation}
\begin{equation} 
h^{ij}\ddot{h}_{ij}+\dot{h}^{ij}\dot{h}_{ij}=0
\end{equation}
Initial conditions are for $h_{i}^{j},\;\dot{h}_{i}^{j}$ with constraints (2.3.4), (2.3.7).
The solution may be obtained via a step by step procedure.

(2.3.4) is the regularity condition.

\section{Radius specified}

\subsection{Observational constraints on the radius}

There are two quantities associated with the radius $R(t)$, the present-day values of which are known from observations: the Hubble parameter
\begin{equation} 
H=\frac{\dot{R}}{R}
\end{equation}
and the deceleration
\begin{equation} 
q=-\frac{R\ddot{R}}{\dot{R}^{2}}=-\frac{\ddot{R}}{RH^{2}}
\end{equation}
The present-day values are [9]
\begin{equation} 
H=(2.28\pm 0.05)\times 10^{-18}\,\mathrm{sec}^{-1}\,,\quad q=-0.595\pm 0.025
\end{equation}
A specified $R(t)$ should be compatible with those values.

\subsection{Radius specification}

Put
\begin{equation} 
R(t)=\sqrt{\varkappa}+(R_{\mathrm{max}}-\sqrt{\varkappa})f(t)
\end{equation}
where
\begin{equation} 
R_{\mathrm{max}}\ggg\sqrt{\varkappa}=t_{\mathrm{Planck}}
\end{equation}
and
\begin{equation} 
f(t+T)=f(t)\;\;\mathrm{for\;some}\;\;T
\end{equation}
\begin{equation} 
f(-t)=f(t),\quad f(0)=0,\quad \dot{f}(0)=0,\quad f(T/2)=1
\end{equation}
We have
\begin{equation} 
f(T/2+\tau)=f(-T/2-\tau)=f(-T/2-\tau+T)=f(T/2-\tau)
\end{equation}
i.e.,
\begin{equation} 
f(T/2-\tau)=f(T/2+\tau),\quad \dot{f}(T/2)=0
\end{equation}
Next, put
\begin{equation} 
f(0)=f_{\mathrm{min}}\,,\quad f(T/2)=f_{\mathrm{max}}
\end{equation}
so that
\begin{equation} 
R(0)=R_{\mathrm{min}}=t_{\mathrm{Planck}}\lll R_{\mathrm{max}}=R(T/2)
\end{equation}

Consider the case
\begin{equation} 
0<t<T/2\;\;\mathrm{and}\;\;R_{\mathrm{max}}f(t)\ggg \sqrt{\varkappa}\,,\;\;
R(t)\approxeq R_{\mathrm{max}}f(t)
\end{equation}

Try the simplest $f$:
\begin{equation} 
f(t)=\sin^{2}\nu t\,,\quad\nu=\pi/T
\end{equation}
We have
\begin{equation} 
R(t)=R_{\mathrm{max}}\sin^{2}\nu t\,,\;\;\dot{R}(t)=R_{\mathrm{max}}2\nu\sin\nu t
\cos\nu t\,,\;\;\ddot{R}(t)=R_{\mathrm{max}}2\nu^{2}(\cos^{2}\nu t-\sin^{2}\nu t)
\end{equation}
and
\begin{equation} 
q=-\frac{1}{2}+\frac{1}{2}\tan^{2}\nu t>-0.5
\end{equation}
Thus, in view of (3.1.3), (3.2.10) is not suitable.

The next candidate is
\begin{equation} 
f(t)=\sin^{4}\nu t\,,\quad \nu=\pi/T
\end{equation}
Now
\begin{equation} 
R(t)=R_{\mathrm{max}}\sin^{4}\nu t\,,\;\;\dot{R}(t)=R_{\mathrm{max}}4\nu\sin^{3}\nu t
\cos\nu t\,,\;\;\ddot{R}(t)=R_{\mathrm{max}}4\nu^{2}\sin^{2}\nu t(3\cos^{2}\nu t-\sin^{2}\nu t)
\end{equation}
and
\begin{equation} 
q=-\frac{3}{4}+\frac{1}{4}\tan^{2}\nu t
\end{equation}
so that
\begin{equation} 
\frac{1}{4}\tan^{2}\nu t=0.75-0.595\pm 0.025=0.155\pm 0.025
\end{equation}
Thus, (3.2.13) is plausible.

\subsection{Parameters determined}

We have two equations:
\begin{equation} 
\tan^{2}\nu t=3+4q
\end{equation}
\begin{equation} 
\frac{4\nu}{\tan\nu t}=H
\end{equation}
Hence
\begin{equation} 
\nu=\frac{1}{4}H\sqrt{3+4q}
\end{equation}
\begin{equation} 
t=\frac{1}{\nu}\tan^{-1}\sqrt{3+4q}\,,\quad t=t_{0}:=t_{\mathrm{present}}
\end{equation}
$T$ is the cosmic period and $\nu$ is the cosmic frequency.

We find
\begin{equation} 
\nu=0.449\times 10^{-18}\,\mathrm{sec}^{-1}\,,\quad T=7.00\times 10^{18}\,\mathrm{sec}=222\times10^{9}\,\mathrm{yr}
\end{equation}

Since $t_{\mathrm{Planck}}$ is a dimensional quantity, there should be another parameter with the same dimensionality---in order that small and large make sense. It is $T$ that is the large counterpart:
\begin{equation} 
t_{\mathrm{Planck}}=5.39\times 10^{-44}\,\mathrm{sec}\lll 7.00\times 10^{18}\,\mathrm{sec}=T
\end{equation}
There are only two constants in the model: the gravitational constant
\begin{equation} 
\varkappa=t_{\mathrm{Planck}}^{2}
\end{equation}
and the cosmic frequency, or
\begin{equation} 
\nu^{2}=(\pi/T)^{2}
\end{equation}
Correspondingly, there is a natural dimensionless constant
\begin{equation} 
\varkappa\nu^{2}=5.86\times 10^{-124}\lll 1
\end{equation}

Put
\begin{equation} 
R_{\mathrm{max}}=\frac{R_{\mathrm{min}}}{\varkappa\nu^{2}}=
\frac{\sqrt{\varkappa}}{\varkappa\nu^{2}}=\frac{1}{\nu(\sqrt{\varkappa}\nu)}
\end{equation}

We obtain
\begin{equation} 
R_{\mathrm{max}}=0.920\times 10^{80}\,\mathrm{sec}
\end{equation}
and
\begin{equation} 
t_{0}:=t_{\mathrm{present}}=1.49\times 10^{18}\,\mathrm{sec}=47.2\times 10^{9}\,\mathrm{yr},
\qquad R_{0}:=R_{\mathrm{present}}=0.136\times 10^{80}\,\mathrm{sec}
\end{equation}
(Without regard to the compenson, $t_{0}=13,7\times 10^{9}\,\mathrm{yr}\;[9].$)

Finally, introduce the inflection instant:
\begin{equation} 
t_{\mathrm{inflection}}\,,\quad \ddot{R}(t_{\mathrm{inflection}})=0
\end{equation}
so that
\begin{equation} 
\ddot{R}(t)\left\{
\begin{array} {lcl}
>0\;\;\mathrm{for}\;\;-t_{\mathrm{inflection}}<t<t_{\mathrm{inflection}}\\
<0\;\;\mathrm{for}\;\;t_{\mathrm{inflection}}<t<T-t_{\mathrm{inflection}}\\
\end{array}
\right.
\end{equation}
and
\begin{equation} 
q(t_{\mathrm{inflection}})=0
\end{equation}
We find
\begin{equation} 
t_{\mathrm{inflection}}=2.34\times 10^{18}\,\mathrm{sec}=74.1\times 10^{9}\,\mathrm{yr}
\end{equation}

So,
\begin{eqnarray}
R_{\mathrm{min}}\leftrightarrow t=0\nonumber\\
<t_{\mathrm{present}}=1.49\times 10^{18}\,\mathrm{sec}=
47.2\times 10^{9}\,\mathrm{yr}\nonumber\\
<t_{\mathrm{inflection}}=2.34\times 10^{18}\,\mathrm{sec}=74.1\times 10^{9}\,\mathrm{yr}\nonumber\\
<T/2=3.50\times 10^{18}\,\mathrm{sec}=111\times 10^{9}\,\mathrm{yr}\leftrightarrow R_{\mathrm{max}}
\end{eqnarray}

\subsection{The first ten minutes}

Put
\begin{equation} 
t=10\,\mathrm{minutes}=600\,\mathrm{sec}
\end{equation}
Since
\begin{equation} 
\nu t\lll 1
\end{equation}
we have
\begin{equation} 
\sin^{4}\nu t=(\nu t)^{4}=52.7\times 10^{-64}
\end{equation}
and
\begin{equation} 
R(t)=R_{\mathrm{max}}(\nu t)^{4}=\frac{1}{\nu(\sqrt{\varkappa}\nu)}(\nu t)^{4}=
0.218\frac{1}{\nu}
\end{equation}
Thus,
\begin{equation} 
\frac{R(10\,\mathrm{minutes})}{R(0)}=\frac{R(10\,\mathrm{minutes})}{R_{\mathrm{min}}}=
\frac{0.218}{\nu}\diagup t_{\mathrm{Planck}}=\frac{0.218}{\sqrt{\varkappa\nu^{2}}}=
0.901\times 10^{61}
\end{equation}

\section{The regular cycling universe}

\subsection{The aging problem}

Although $R(t)$ is a periodic function, the construction of a cycling model is not yet complete. Such a model faces the aging problem [10,11]. We quote Weinberg [11]:

``Some cosmologists are philosophically attracted to the oscillating model, especially because \ldots it nicely avoids the problem of Genesis. It does, however, face one severe theoretical difficulty. In each cycle the ratio of photons to nuclear particles\ldots is slightly increased by a kind of friction (known as ``bulk viscosity'') as the universe expands and contracts. As far as we know, the universe would then start each new cycle with a new, slightly larger ratio of photons to nuclear particles. Right now this ratio is large, but not infinite, so it is hard to see how the universe could have previously experienced an infinite number of cycles.''

\subsection{Big Jump}

A feasible resolution of the aging problem is via the following quantum jump:
\begin{equation} 
\mathrm{at}\;\;t=0\quad \Psi(t-0)\rightarrow\Psi(t+0)=\Psi_{0},\quad\Psi_{0}\;\,\mathrm{preassigned} 
\end{equation}
which may be called the Big Jump.

\subsection{The regular cycling universe}

Now we have a regular cycling model of the universe. Specifically,
\begin{equation} 
R(t)=R_{\mathrm{min}}=\sqrt{\varkappa}\quad
\mathrm{at}\;\; t=nT,\;\;n=\cdots,-2,-1,0,1,2,\cdots
\end{equation}
\begin{equation} 
R(t)=R_{\mathrm{max}}=\frac{1}{\nu\sqrt{\varkappa\nu^{2}}}\quad \mathrm{at}
\;\;t=(n+1/2)T,\;\;n=\cdots,-2,-1,0,1,2,\cdots
\end{equation}
with
\begin{equation} 
\mathrm{at}\;\;t=nT\quad \Psi(nT-0)\rightarrow\Psi(nT+0)=\Psi_{0},\quad\Psi_{0}\;\,\mathrm{preassigned}
\end{equation}
It must be emphasized that, in our treatment,
\begin{equation} 
\mathrm{regularity}\Rightarrow \mathrm{cyclicity}
\end{equation}
but, on the other hand,
\begin{equation} 
\mathrm{cyclicity}\nRightarrow \mathrm{regularity}
\end{equation}
So, the closed Friedmann model is oscillating but not regular (see [12] on cycling models).

\subsection{Pseudo-black hole}

Regularity implies the absence of black holes. But a contracting-expanding (or pulsating) massive star may simulate the behavior of a black hole. We quote Straumann [5]:

``An external observer far away from the star will see it reach the horizon only after an infinitely long time. As a result of the gravitational time dilation, the star ``freezes'' at the Schwarzschild horizon. However, in practice, the star will suddenly become invisible, since the redshift will start to increase exponentially\ldots, and the luminosity decreases correspondingly.''

A pure compenson object may appear to be a black hole as well.

\section{The isotropic universe}

\subsection{Metric and measure}

Metric of the closed isotropic universe is of the form [8]
\begin{equation} 
\mathrm{d}s^{2}=
\mathrm{d}t^{2}-R^{2}(t)[\mathrm{d}\chi^{2}+
\sin^{2}\chi(\mathrm{d}\theta^{2}+\sin^{2}\theta\,\mathrm{d}\varphi^{2})]\,,
\quad 0\leq \chi\leq\pi,\;0\leq\theta\leq\pi,\;0\leq\varphi\leq 2\pi
\end{equation}
and the measure associated with it is (2.3.3):
\begin{equation} 
\mathrm{d}\mu(\omega)=
\sin^{2}\chi\sin\theta\,\mathrm{d}\chi\mathrm{d}\theta\mathrm{d}\varphi\,,
\quad\omega=(\chi,\theta,\varphi)
\end{equation}
Thus, equations (2.3.4), (2.3.5), (2.3.7), (2.3.9) hold identically.

\subsection{The extended Einstein equations}

The extended Einstein equations are
\begin{equation} 
2\frac{\ddot{R}}{R}+\frac{\dot{R}^{2}}{R^{2}}+\frac{1}{R^{2}}=
8\pi\varkappa(T_{1}^{1}+\Theta^{1}_{1})
\end{equation}
\begin{equation} 
3\left(\frac{\dot{R}^{2}}{R^{2}}+\frac{1}{R^{2}}\right)=
8\pi\varkappa(T_{0}^{0}+\Theta_{0}^{0})
\end{equation}
where
\begin{equation} 
R(t)=\sqrt{\varkappa}+(R_{\mathrm{max}}-\sqrt{\varkappa})\sin^{4}\nu t
\end{equation}

\subsection{Dynamical equation. Compenson pressure}

There is only one dynamical equation (5.2.1), which determines the compenson pressure,
\begin{equation} 
p_{\mathrm{compenson}}:=-\Theta^{1}_{1}
\end{equation}
\begin{equation} 
p_{\mathrm{compenson}}=-\frac{1}{8\pi\varkappa}
\left(2\frac{\ddot{R}}{R}+\frac{\dot{R}^{2}}{R^{2}}+\frac{1}{R^{2}}\right)
-p_{\mathrm{matter}}
\end{equation}
where the matter pressure
\begin{equation} 
p_{\mathrm{matter}}=-T_{1}^{1}
\end{equation}

\subsection{Compenson density}

Equation (5.2.2), or
\begin{equation} 
3\left(\frac{\dot{R}^{2}}{R^{2}}+\frac{1}{R^{2}}\right)=
8\pi\varkappa(\varrho_{\mathrm{matter}}+\varrho_{\mathrm{compenson}})
\end{equation}
\begin{equation} 
\varrho_{\mathrm{matter}}=T_{0}^{0}\,,\quad \varrho_{\mathrm{compenson}}=\Theta_{0}^{0}
\end{equation}
determines the compenson density:
\begin{equation} 
\varrho_{\mathrm{compenson}}=\frac{3}{8\pi\varkappa}
\left(\frac{\dot{R}^{2}}{R^{2}}+\frac{1}{R^{2}}\right)-\varrho_{\mathrm{matter}}
\end{equation}

\subsection{Matter density}

We have
\begin{equation} 
\varrho_{\mathrm{dark\,energy}}+\varrho_{\mathrm{dark\,matter}}+
\varrho_{\mathrm{matter}}\leftrightarrow 1
\end{equation}
and [9]
\begin{equation} 
\mathrm{at}\;\;t_{0}=t_{\mathrm{present}}\quad \varrho_{\mathrm{dark\,energy}}=
\frac{\Lambda}{8\pi\varkappa}\leftrightarrow 0.732\,,\;\;\varrho_{\mathrm{dark\,matter}}
\leftrightarrow 0.223\,,\;\;\varrho_{\mathrm{matter}}\leftrightarrow 0.044
\end{equation}
with
\begin{equation} 
\Lambda=0.732\times 3H_{0}^{2}
\end{equation}
Thus,
\begin{equation} 
\mathrm{at}\;\;t_{0}=t_{\mathrm{present}}\quad \varrho_{\mathrm{matter}}=
0.044\times \frac{\varrho_{\mathrm{dark\,energy}}}{0.732}=
\frac{3\times 0.044}{8\pi}\frac{H_{0}^{2}}{\varkappa}
\end{equation}
By (3.3.3),
\begin{equation} 
H_{0}=\frac{\nu}{0.197}
\end{equation}
so that
\begin{equation} 
\mathrm{at}\;\;t_{0}=t_{\mathrm{present}}\quad \varrho_{\mathrm{matter}}=
\frac{0.044}{0.0388}\times\frac{3}{8\pi}\frac{\nu^{2}}{\varkappa}
\end{equation}

Next,
\begin{equation} 
\varrho_{\mathrm{matter}}(0)\gtrapprox \varrho_{\mathrm{matter}}(t_{0})
\left(\frac{R_{\mathrm{max}}}{\sqrt{\varkappa}}\right)^{3}=
\varrho_{\mathrm{matter}}(t_{0})\frac{1}{(\varkappa\nu^{2})^{3}}\approx
 \frac{3}{8\pi}\frac{1}{\varkappa^{2}}\frac{1}{(\varkappa\nu^{2})^{2}}
 =\frac{3}{8\pi}\frac{\varrho_{\mathrm{Planck}}}{(\varkappa\nu^{2})^{2}}
\end{equation}
where
\begin{equation} 
\varrho_{\mathrm{Planck}}=\frac{1}{\varkappa^{2}}=\frac{1}{t_{\mathrm{Planck}}^{4}}
\end{equation}
Thus,
\begin{equation} 
\varrho_{\mathrm{matter}}(0)\sim \frac{\varrho_{\mathrm{Planck}}}
{(\varkappa\nu^{2})^{2}}
\end{equation}

\subsection{Density parameters}

Equation (1.4.3) reduces to
\begin{equation} 
3\left(\frac{\dot{R}^{2}}{R^{2}}+\frac{1}{R^{2}}\right)=
8\pi\varkappa(\varrho_{\mathrm{matter}}+\varrho_{\mathrm{dark\,matter}}+
\varrho_{\mathrm{dark\,energy}})
\end{equation}
or
\begin{equation} 
H^{2}+\frac{1}{R^{2}}=\frac{8\pi\varkappa}{3}
(\varrho_{\mathrm{matter}}+\varrho_{\mathrm{dark\,matter}}+
\varrho_{\mathrm{dark\,energy}})
\end{equation}
Now, the critical density [9] is
\begin{equation} 
\varrho_{\mathrm{critical}}:=\frac{3H^{2}}{8\pi\varkappa}
\end{equation}
So,
\begin{equation} 
\frac{1}{\varrho_{\mathrm{critical}}}
(\varrho_{\mathrm{matter}}+\varrho_{\mathrm{dark\,matter}}+
\varrho_{\mathrm{dark\,energy}})=1+\frac{1}{\dot{R}^{2}}
\end{equation}
or, in terms of density parameter [9],
\begin{equation} 
\Omega_{...}:=\frac{\varrho_{...}}{\varrho_{\mathrm{critical}}}
\end{equation}
\begin{equation} 
\Omega=\Omega_{\mathrm{matter}}+\Omega_{\mathrm{dark\,matter}}+
\Omega_{\mathrm{dark\,energy}}=1+\frac{1}{\dot{R}^{2}}
\end{equation}

Again,
\begin{equation} 
\dot{R}=\frac{4}{\sqrt{\varkappa}\nu}\sin^{3}\nu t\cos\nu t
\end{equation}
so that
\begin{equation} 
\Omega=1+\frac{\varkappa\nu^{2}}{16\sin^{6}\nu t\cos^{2}\nu t}=
1+\frac{(t_{\mathrm{Planck}}\nu)^{2}}{16\sin^{6}\nu t\cos^{2}\nu t}=
1+\frac{0.366\times 10^{-124}}{\sin^{6}\nu t\cos^{2}\nu t}
\end{equation}

\section{On justification of the concept of compenson}

\subsection{Energy-momentum compenson}

The energy-momentum compenson $\Theta_{\mu}^{0}$ compensates for quantum jumps of $T_{\mu}^{0}$ in (1.3.3), (1.3.4).

\subsection{Stress compenson}

We have (2.1.1):
\begin{equation} 
g_{ij}(t,\underline{x})=-R^{2}(t)h_{ij}(t,\underline{x})\quad \mathrm{with}\;\; R(t)\;\mathrm{preassigned}
\end{equation}
In order that this make sense, i.e., be nontrivial, there should be an additional equation (aside from the Einstein equations) for $h_{ij}$ and, accordingly, a stress compenson function compensating for the equation.

\section{Stress compenson specified}

\subsection{The spatial Einstein equation with compenson}

The spatial Einstein equation with compenson is
\begin{equation} 
G^{j}_{i}=8\pi\varkappa(T_{i}^{j}+\Theta_{i}^{j})
\end{equation}

\subsection{Pressure compenson}

We shall accept the simplest form of the stress compenson:
\begin{equation} 
\Theta_{i}^{j}=\vartheta\,\delta^{j}_{i}
\end{equation}
the function (i.e., scalar) $\vartheta$ will be called the pressure compenson:
\begin{equation} 
\vartheta=-p_{\mathrm{compenson}}
\end{equation}

\section{Pressure compenson eliminated}

\subsection{The extended spatial Einstein equation}

The extended spatial Einstein equation takes the form
\begin{equation} 
G^{j}_{i}=8\pi\varkappa(T_{i}^{j}+\vartheta\,\delta_{i}^{j})
\end{equation}
This is a system of 6 equations for metric components and the pressure compenson.

\subsection{Eliminating pressure compenson}

Let us express the Einstein tensor in (8.1.1) through the Ricci tensor:
\begin{equation} 
R_{i}^{j}-\frac{1}{2}R_{\mu}^{\mu}\,\delta_{i}^{j}=8\pi\varkappa\, T_{i}^{j}+8\pi\varkappa\,\vartheta\,\delta_{i}^{j}
\end{equation}
We have
\begin{equation} 
R_{l}^{l}-\frac{3}{2}R_{\mu}^{\mu}=8\pi\varkappa\,T_{l}^{l}+24\pi\varkappa\,\vartheta
\end{equation}
whence
\begin{equation} 
\vartheta=\frac{1}{24\pi\varkappa}\left(R_{l}^{l}-\frac{3}{2}R_{\mu}^{\mu}\right)-\frac{1}{3}T_{l}^{l}
\end{equation}
Substitute (8.2.3) into (8.2.1):
\begin{equation} 
R_{i}^{j}-\frac{1}{3}R_{l}^{l}\,\delta_{i}^{j}=8\pi\varkappa\left(T_{i}^{j}-
\frac{1}{3}T_{l}^{l}\delta_{i}^{j}\right)
\end{equation}

Let us introduce a traceless tensor
\begin{equation} 
\bar{X}_{i}^{j}:=X_{i}^{j}-\frac{1}{3}X_{l}^{l}\,\delta_{i}^{j}\,,\quad\bar{X}_{l}^{l}=0
\end{equation}
So we have obtained the equation
\begin{equation} 
\bar{R}_{i}^{j}=8\pi\varkappa\,\bar{T}_{i}^{j}
\end{equation}
which represents a system of 5 equations for metric components.

\section{Metric and the system of dynamical equations}

\subsection{Metric}

Metric is of the form
\begin{equation} 
\mathrm{d}s^{2}=\mathrm{d}x^{0\,2}+g_{ij}\mathrm{d}x^{i}\mathrm{d}x^{j}
\end{equation}
It is convenient to represent metric also in the form
\begin{equation} 
\mathrm{d}s^{2}=\mathrm{d}t^{2}-\mathrm{d}l^{2}
\end{equation}
with
\begin{equation} 
\mathrm{d}l^{2}=
R^{2}(t)h_{ij}(\underline{x},t)\mathrm{d}x^{i}\mathrm{d}x^{j}
\end{equation}

\subsection{The complete system of dynamical equations for metric}

We have (2.3.9)
\begin{equation} 
h^{ij}\ddot{h}_{ij}+\dot{h}^{ij}\dot{h}_{ij}=0
\end{equation}
Again,
\begin{equation} 
h^{il}h_{lj}=\delta_{j}^{i}\,,\quad \dot{h}^{il}h_{lj}+h^{il}\dot{h}_{lj}=0
\end{equation}
whence
\begin{equation} 
\dot{h}^{ij}=-h^{il}\dot{h}_{lk}h^{kj}
\end{equation}
and, in view of (2.3.7),
\begin{equation} 
\dot{h}^{ij}h_{ij}=0
\end{equation}
Equation (9.2.1) takes the form
\begin{equation} 
h^{ij}\ddot{h}_{ij}-h^{il}\dot{h}_{lk}h^{kj}\dot{h}_{ij}=0
\end{equation}

Equations (8.2.6)
\begin{equation} 
\bar{R}_{i}^{j}=8\pi\varkappa\,\bar{T}_{i}^{j}
\end{equation}
and (9.2.5) form a complete system of 6 dynamical equations for 6 metric components $h_{ij}$.

\section{The Christoffel symbols and Ricci tensor}

\subsection{The Christoffel symbols}

Let us introduce
\begin{equation} 
\beta_{ij}:=R^{2}h_{ij}\,,\quad \beta^{ij}=\frac{1}{R^{2}}h^{ij}\,,\quad
\beta^{il}\beta_{lj}=h^{il}h_{lj}=\delta_{j}^{i}
\end{equation}
and
\begin{equation} 
\sigma_{ij}:=\frac{\partial}{\partial t}\beta_{ij}=\dot{\beta}_{ij}
\end{equation}

The Christoffel symbols are (here and further we exploit [8])
\begin{eqnarray}
\Gamma_{00}^{0}=0\,,\quad\Gamma_{00}^{i}=0\,,\quad\Gamma_{0i}^{0}=0\nonumber\\
\Gamma_{ij}^{0}=\frac{1}{2}\sigma_{ij}\,,\quad \Gamma_{0j}^{i}=\frac{1}{2}\sigma_{j}^{i}\nonumber\\
\Gamma_{jl}^{i}=\lambda_{jl}^{i}
\end{eqnarray}
where
\begin{equation} 
\sigma_{j}^{i}=\beta^{il}\sigma_{lj}=h^{il}\dot{h}_{lj}+2\frac{\dot{R}}{R}\delta_{j}^{i}
\end{equation}
and
\begin{equation} 
\lambda_{jl}^{i}=\frac{1}{2}\beta^{im}(\beta_{ml,j}+\beta_{jm,l}-\beta_{jl,m})
=\frac{1}{2}h^{im}(h_{ml,j}+h_{jm,l}-h_{jl,m})
\end{equation}

\subsection{The Ricci tensor}

The Ricci tensor components are these:
\begin{equation} 
R_{0}^{0}=R_{00}=-\frac{1}{2}\frac{\partial}{\partial t}\sigma_{l}^{l}-
\frac{1}{4}\sigma_{l}^{k}\sigma_{k}^{l}
\end{equation}
\begin{equation} 
R_{i}^{0}=R_{0i}=\frac{1}{2}(\sigma_{i;l}^{l}-\sigma_{l;i}^{l})
\end{equation}
\begin{equation} 
R_{ij}=\frac{1}{2}\frac{\partial}{\partial t}\sigma_{ij}+\frac{1}{4}
(\sigma_{ij}\sigma_{l}^{l}-2\sigma_{i}^{l}\sigma_{jl})+Q_{ij}
\end{equation}
\begin{equation} 
R_{i}^{j}=-\frac{1}{2\sqrt{|\beta|}}\frac{\partial}{\partial t}(\sqrt{|\beta|}\sigma_{i}^{j})-
Q_{i}^{j}
\end{equation}
Here
\begin{equation} 
Q_{ji}=Q_{ij}=\frac{\partial\Gamma_{ij}^{l}}{\partial x^{l}}-\frac{\partial\Gamma_{il}^{l}}{\partial x^{j}}
+\Gamma_{ij}^{l}\Gamma_{lm}^{m}-\Gamma_{il}^{m}\Gamma_{jm}^{l}
\end{equation}
\begin{equation} 
Q_{i}^{j}=\beta^{jk}Q_{ki}=\frac{1}{R^{2}}h^{jk}Q_{ki}
\end{equation}

Next,
\begin{equation} 
\sigma_{l}^{l}=h^{lk}\dot{h}_{lk}+2\frac{\dot{R}}{R}\delta_{l}^{l}=6\frac{\dot{R}}{R}
\end{equation}
whence
\begin{equation} 
\sigma_{l;i}^{l}=\sigma_{l,i}^{l}=0
\end{equation}
\begin{equation} 
\frac{\partial}{\partial t}\sigma_{l}^{l}=6\left(\frac{\ddot{R}}{R}-
\frac{\dot{R}^{2}}{R^{2}}\right)
\end{equation}
We obtain
\begin{equation} 
\sigma_{l}^{k}\sigma_{k}^{l}=k^{km}\dot{h}_{ml}h^{ln}\dot{h}_{nk}+
12\frac{\dot{R}^{2}}{R^{2}}
\end{equation}
and
\begin{equation} 
\sigma_{i;l}^{l}=(h^{lk}\dot{h}_{ki})_{,l}-\Gamma_{il}^{m}h^{lk}\dot{h}_{km}+
\Gamma_{ml}^{l}h^{mk}\dot{h}_{ki}
\end{equation}

Again, by (10.1.1)
\begin{equation} 
|\beta|=R^{6}|h|
\end{equation}
so that, in view of (2.3.5), (10.2.4) reduces to
\begin{equation} 
R_{i}^{j}=-\frac{1}{2R^{3}}\frac{\partial}{\partial t}(R^{3}\sigma_{i}^{j})-Q_{i}^{j}
\end{equation}

\section{The metrodynamical equation. \\The Haar measure condition as an integral of motion}

\subsection{The metrodynamical equation}

Let us return to (9.2.6):
\begin{equation} 
\bar{R}_{i}^{j}=8\pi\varkappa\,\bar{T}_{i}^{j}
\end{equation}
By (10.2.13)
\begin{equation} 
\bar{R}_{i}^{j}=-\frac{1}{2R^{3}}\frac{\partial}{\partial t}(R^{3}\bar{\sigma}_{i}^{j})
-\bar{Q}_{i}^{j}
\end{equation}
Again,
\begin{equation} 
\bar{\sigma}_{i}^{j}=\sigma_{i}^{j}-\frac{1}{3}\sigma_{l}^{l}\delta_{i}^{j}=
h^{jk}\dot{h}_{ki}
\end{equation}
so that
\begin{equation} 
\bar{R}_{i}^{j}=-\frac{3}{2}\frac{\dot{R}}{R}(h^{jk}\dot{h}_{ki})-
\frac{1}{2}(\dot{h}^{jk}\dot{h}_{ki}+h^{jk}\ddot{h}_{ki})-\bar{Q}_{i}^{j}
\end{equation}
Next, substituting (9.2.3) for $\dot{h}^{jk}$, we represent (11.1.1) as
\begin{equation} 
h^{jk}\left[\frac{1}{2}\ddot{h}_{ki}-\frac{1}{2}\dot{h}_{km}h^{ml}\dot{h}_{li}+
\frac{3}{2}\frac{\dot{R}}{R}\dot{h}_{ki}\right]+\bar{Q}_{i}^{j}=-8\pi\varkappa\bar{T}_{i}^{j}
\end{equation}
or
\begin{equation} 
\ddot{h}_{ij}+3\frac{\dot{R}}{R}\dot{h}_{ij}-\dot{h}_{im}h^{mn}\dot{h}_{nj}
+2h_{il}\bar{Q}_{j}^{l}=-16\pi\varkappa h_{il}\bar{T}_{j}^{l}
\end{equation}

Now, let us return to (8.2.5)
\begin{equation} 
\bar{X}_{i}^{j}:=X_{i}^{j}-\frac{1}{3}X_{l}^{l}\delta_{i}^{j}\,,\quad\bar{X}_{l}^{l}=0
\end{equation}

and introduce
\begin{equation} 
\bar{X}_{ij}:=X_{ij}-\frac{1}{3}h_{ij}h^{mn}X_{mn}\,,\quad
X_{ij}=X_{ji}\,,\quad h^{ij}\bar{X}_{ij}=0
\end{equation}
We find
\begin{equation} 
h_{il}\bar{Q}_{j}^{l}=\frac{1}{R^{2}}\beta_{il}\bar{Q}_{j}^{l}=\frac{1}{R^{2}}\bar{Q}_{ij}
\end{equation}
and (see [8])
\begin{equation} 
h_{il}\bar{T}_{j}^{l}=\frac{1}{R^{2}}\beta_{il}\bar{T}_{j}^{l}
=-\frac{1}{R^{2}}g_{il}\bar{T}_{j}^{l}
=-\frac{1}{R^{2}}\bar{T}_{ij}
\end{equation}
Thus, (11.1.6) takes a resultant form
\begin{equation} 
\ddot{h}_{ij}+3\frac{\dot{R}}{R}\dot{h}_{ij}-\dot{h}_{im}h^{mn}\dot{h}_{nj}
+\frac{2}{R^{2}}\bar{Q}_{ij}=16\pi\varkappa\frac{1}{R^{2}}\bar{T}_{ij}
\end{equation}
This is the metrodynamical equation---a dynamical equation for metric components.

\subsection{The Haar measure condition as an integral of motion of the metrodynamical equation}

From the metrodynamical equation follows
\begin{equation} 
h^{ji}\ddot{h}_{ij}-h^{ji}\dot{h}_{im}h^{mn}\dot{h}_{nj}+
3\frac{\dot{R}}{R}(h^{ji}\dot{h}_{ij})=0
\end{equation}
or, by using (9.2.3),
\begin{equation} 
h^{ji}\ddot{h}_{ij}+\dot{h}^{ji}\dot{h}_{ij}+
3\frac{\dot{R}}{R}(h^{ji}\dot{h}_{ij})=0
\end{equation}
i.e.,
\begin{equation} 
\frac{\partial}{\partial t}(h^{ji}\dot{h}_{ij})+3\frac{\dot{R}}{R}(h^{ji}\dot{h}_{ij})=0
\end{equation}

By the Haar measure condition (2.3.7)
\begin{equation} 
h^{ji}\dot{h}_{ij}=0
\end{equation}
so that (11.2.3) is an identity. However, we may reverse the treatment, i.e., treat (11.2.3) as an equation. Then we obtain
\begin{equation} 
[\partial(h^{ji}\dot{h}_{ij})/\partial t]/(h^{ji}\dot{h}_{ij})+
3[\mathrm{d}R/\mathrm{d}t]/R=0
\end{equation}
i.e.,
\begin{equation} 
\frac{\partial}{\partial t}\ln[(h^{ji}\dot{h}_{ij})R^{3}]=0
\end{equation}
whence
\begin{equation} 
h^{ji}\dot{h}_{ij}=\frac{f(\underline{x})}{R^{3}(t)}
\end{equation}
Equation (11.2.4) implies
\begin{equation} 
f(\underline{x})=0
\end{equation}
Thus, the Haar measure condition may be regarded as an integral of motion for the metrodynamical equation.

\section{Compenson determined}

\subsection{Pressure compenson}

The pressure compenson is (8.2.3)
\begin{equation} 
\vartheta=\frac{1}{24\pi\varkappa}\left(R_{l}^{l}-\frac{3}{2}R_{\mu}^{\mu}\right)-
\frac{1}{3}T_{l}^{l}=
-\frac{1}{24\pi\varkappa}\left[\frac{1}{2}(R_{l}^{l}+3R_{0}^{0})+
8\pi\varkappa T_{l}^{l}\right]
\end{equation}
By (10.2.13)
\begin{equation} 
R_{l}^{l}=-\frac{1}{2R^{3}}\frac{\partial}{\partial t}(R^{3}\sigma_{l}^{l})-Q_{l}^{l}
\end{equation}
and by (10.2.6)
\begin{equation} 
Q_{l}^{l}=\frac{1}{R^{2}}h^{lk}Q_{kl}
\end{equation}
Using (10.2.1), (10.2.7), (10.2.9) we obtain
\begin{equation} 
\vartheta=\frac{1}{8\pi\varkappa}\left\{\left[2\frac{\ddot{R}}{R}
+\frac{\dot{R}^{2}}{R^{2}}+\frac{1}{6R^{2}}h^{lk}Q_{kl}+
\frac{1}{8}h^{lm}\dot{h}_{mk}h^{kn}\dot{h}_{nl}\right]-\frac{8\pi\varkappa}{3}T_{l}^{l}\right\}
\end{equation}
Notation:
\begin{equation} 
(\dot{h})_{nl}:=\dot{h}_{nl}\,,\quad (\dot{h})^{ln}:=h^{lm}(\dot{h})_{mn}h^{kn}\,,\quad
(\dot{h})^{ln}\neq \dot{h}^{ln}
\end{equation}
(12.1.4) takes the form
\begin{equation} 
\vartheta=\frac{1}{8\pi\varkappa}\left\{\left[2\frac{\ddot{R}}{R}
+\frac{\dot{R}^{2}}{R^{2}}+\frac{1}{6R^{2}}h^{lk}Q_{kl}+
\frac{1}{8}(\dot{h})^{ln}(\dot{h})_{nl}\right]-\frac{8\pi\varkappa}{3}T_{l}^{l}\right\}
\end{equation}

\subsection{Energy compenson}

From the extended Einstein equation (1.3.3)
\begin{equation} 
R_{0}^{0}-\frac{1}{2}R_{\mu}^{\mu}=8\pi\varkappa(T_{0}^{0}+\Theta_{0}^{0})
\end{equation}
follows the expression for the energy compenson
\begin{equation} 
\Theta_{0}^{0}=\frac{1}{16\pi\varkappa}(R_{0}^{0}-R_{l}^{l})-T_{0}^{0}
\end{equation}
We obtain
\begin{equation} 
\Theta_{0}^{0}=\frac{1}{8\pi\varkappa}\left\{\left[3\frac{\dot{R}^{2}}{R^{2}}+
\frac{1}{2R^{2}}h^{lk}Q_{kl}-\frac{1}{8}(\dot{h})^{lk}(\dot{h})_{kl}\right]
-8\pi\varkappa T_{0}^{0}\right\}
\end{equation}

\subsection{Momentum compenson}

From the extended Einstein equation (1.3.4)
\begin{equation} 
R_{i}^{0}=8\pi\varkappa(T_{i}^{0}+\Theta_{i}^{0})
\end{equation}
follows the expression for the momentum compenson
\begin{equation} 
\Theta_{i}^{0}=\frac{1}{8\pi\varkappa}R_{i}^{0}-T_{i}^{0}
\end{equation}
By (10.2.2), (10.2.8), (10.2.11)
\begin{equation} 
R_{i}^{0}=\frac{1}{2}\left[(h^{lk}\dot{h}_{ki})_{,l}-
\Gamma_{il}^{m}h^{lk}\dot{h}_{km}+\Gamma_{ml}^{l}h^{mk}\dot{h}_{ki}\right]
\end{equation}
Notation:
\begin{equation} 
(\dot{h})_{i}^{j}:=h^{jk}(\dot{h})_{ki}\,,\quad (\dot{h})^{j}_{i}\neq \dot{h}^{j}_{i}=0
\end{equation}
Thus,
\begin{equation} 
\Theta_{i}^{0}=\frac{1}{8\pi\varkappa}\left\{\frac{1}{2}\left[(\dot{h})_{i,l}^{l}-
\Gamma_{il}^{m}(\dot{h})_{m}^{l}+\Gamma_{ml}^{l}(\dot{h})_{i}^{m}\right]
-8\pi\varkappa T_{i}^{0}\right\}
\end{equation}

\subsection{Quantum jumps and compenson}

From (12.1.6), (12.2.3), (12.3.5) it follows that under quantum jumps of the energy-momentum tensor
\begin{equation} 
T_{\mu}^{\nu}=(\Psi, \hat{T}_{\mu}^{\nu}\Psi)
\end{equation}
the quantities
\begin{equation} 
\frac{1}{3}T_{l}^{l}+\theta\,,\quad T_{0}^{0}+\Theta_{0}^{0}\,,\quad T_{i}^{0}+\Theta_{i}^{0}
\end{equation}
remain continuous.

Next, let us return to (9.2.1)
\begin{equation} 
h^{ij}\ddot{h}_{ij}+\dot{h}^{ij}\dot{h}_{ij}=0
\end{equation}
Since
\begin{equation} 
h^{ij}\ddot{h}_{ij}+\dot{h}^{ij}\dot{h}_{ij}=\frac{\partial}{\partial t}(h^{ij}\dot{h}_{ij})
\end{equation}
and according to (11.2.4)
\begin{equation} 
h^{ij}\dot{h}_{ij}=0
\end{equation}
the quantity
\begin{equation} 
h^{ij}\ddot{h}_{ij}
\end{equation}
is continuous under quantum jumps.

\section{The isotropic universe. Background and deviation}

\subsection{The isotropic universe}

Let us return to the isotropic universe, in which it holds that
\begin{equation} 
A_{i}^{0}=0\,,\quad A_{i}^{j}=0\;\;\mathrm{for}\;i\neq j\,,
\quad A_{1}^{1}=A_{2}^{2}=A_{3}^{3}
\end{equation}
From (11.1.7) follows
\begin{equation} 
\bar{A}_{i}^{j}=0
\end{equation}
and by (11.1.9), (11.1.10)
\begin{equation} 
\bar{T}_{ij}=0\,,\quad \bar{Q}_{ij}=0
\end{equation}

The extended Einstein equations are
\begin{equation} 
2\frac{\ddot{R}}{R}+\frac{\dot{R}^{2}}{R^{2}}+\frac{1}{R^{2}}=8\pi\varkappa(T_{1}^{1}+\vartheta)
\end{equation}
\begin{equation} 
3\left(\frac{\dot{R}^{2}}{R^{2}}+\frac{1}{R^{2}}\right)=8\pi\varkappa(T_{0}^{0}+\Theta_{0}^{0})
\end{equation}

The Haar measure condition is satisfied identically. The metrodynamical equation in the form of (11.1.6) reduces to an identity.

The expression for the pressure compenson (12.1.6) reduces to
\begin{equation} 
\vartheta=\frac{1}{8\pi\varkappa}\left\{\left[2\frac{\ddot{R}}{R}+
\frac{\dot{R}^{2}}{R^{2}}+\frac{1}{6R^{2}}h^{mn}Q_{mn}\right]-
\frac{8\pi\varkappa}{3}T_{l}^{l}\right\}
\end{equation}

From (13.1.6), (13.1.4) follows
\begin{equation} 
h^{mn}Q_{mn}=6
\end{equation}
and by (12.1.3)
\begin{equation} 
Q_{l}^{l}=\frac{6}{R^{2}}
\end{equation}

\subsection{Background and deviation}

Let us introduce the notion of a background and deviation with the background corresponding to the isotropic universe. So,
\begin{equation} 
(T+\Theta)=(T+\Theta)_{\mathrm{background}}+(T+\Theta)_{\mathrm{deviation}}
\end{equation}
\begin{equation} 
8\pi\varkappa(T_{1}^{1}+\vartheta)_{\mathrm{background}}=2\frac{\ddot{R}}{R}+
\frac{\dot{R}^{2}}{R^{2}}+\frac{1}{R^{2}}
\end{equation}
\begin{equation} 
8\pi\varkappa(T_{0}^{0}+\Theta_{0}^{0})_{\mathrm{background}}=
3\left(\frac{\dot{R}^{2}}{R^{2}}+\frac{1}{R^{2}}\right)
\end{equation}
\begin{eqnarray}
A_{\mathrm{background}\;i}{}^{0}=0,\quad\bar{A}_{\mathrm{background}\,ij}=0\nonumber\\
A_{i}^{0}=A_{\mathrm{deviation}\;i}{}^{0}\,,\quad\bar{A}_{ij}=\bar{A}_{\mathrm{deviation}\,ij}\,,
\qquad A=T,Q
\end{eqnarray}

\subsection{Background eliminated}

The metrodynamical equation (11.1.11) takes the form
\begin{equation} 
\ddot{h}_{ij}+3\frac{\dot{R}}{R}\dot{h}_{ij}-(\dot{h})^{in}(\dot{h})_{nj}
+\frac{2}{R^{2}}\bar{Q}_{\mathrm{deviation}\;ij}=
16\pi\varkappa\frac{1}{R^{2}}\bar{T}_{\mathrm{deviation}\;ij}
\end{equation}
We rewrite equation (12.2.3) as
\begin{equation} 
\left[\frac{1}{2}Q_{m}^{m}-\frac{3}{R^{2}}\right]-\frac{1}{8}(\dot{h})^{lk}(\dot{h})_{kl}=
8\pi\varkappa(T+\Theta)_{\mathrm{deviation}}{}\,^{0}_{0}
\end{equation}
and equation (12.3.5) as
\begin{equation} 
\frac{1}{2}\left[(\dot{h})_{i,l}^{l}-\Gamma_{il}^{m}(\dot{h})_{m}^{l}+
\Gamma_{ml}^{l}(\dot{h})_{i}^{m}\right]=8\pi\varkappa(T+\Theta)_{\mathrm{deviation}}\,{}^{0}_{i}
\end{equation}

Next, (12.1.6) may be written as
\begin{equation} 
\vartheta=\frac{1}{8\pi\varkappa}\left\{\left[\left(
\frac{1}{6}Q_{l}^{l}-\frac{1}{R^{2}}\right)+\frac{1}{8}(\dot{h})^{ln}(\dot{h})_{nl}\right]+
\left[\left(2\frac{\ddot{R}}{R}+\frac{\dot{R}^{2}}{R^{2}}+\frac{1}{R^{2}}\right)-
\frac{8\pi\varkappa}{3}T_{l}^{l}\right]\right\}
\end{equation}
or
\begin{equation} 
\vartheta=\frac{1}{8\pi\varkappa}\left\{\left[\left(
\frac{1}{6}Q_{l}^{l}-\frac{1}{R^{2}}\right)+\frac{1}{8}(\dot{h})^{ln}(\dot{h})_{nl}\right]+
8\pi\varkappa\,\theta_{\mathrm{background}}-
\frac{8\pi\varkappa}{3}T_{\mathrm{deviation}}\,{}^{l}_{l}\right\}
\end{equation}
so that
\begin{equation} 
\vartheta_{\mathrm{deviation}}=\frac{1}{8\pi\varkappa}\left\{\left[\left(
\frac{1}{6}Q_{l}^{l}-\frac{1}{R^{2}}\right)+\frac{1}{8}(\dot{h})^{ln}(\dot{h})_{nl}\right]
-\frac{8\pi\varkappa}{3}T_{\mathrm{deviation}}\,{}^{l}_{l}\right\}
\end{equation}

\newpage
\section*{Acknowledgments}

I would like to thank Alex A. Lisyansky for support and Stefan V.
Mashkevich for helpful discussions.

\end{document}